\RequirePackage{fix-cm}
\documentclass{svjour3}                     
\pdfoutput=1

\smartqed  
\usepackage{graphicx}
\usepackage{hyperref}

\usepackage[utf8]{inputenc}

\usepackage{amsfonts,amsmath,amssymb}
\usepackage{xcolor}
\usepackage{graphicx}
\usepackage{mdframed}

\spnewtheorem{thm}{Theorem}{\bfseries }{\itshape }
\spnewtheorem{prop}[thm]{Proposition}{\bfseries }{\itshape }
\spnewtheorem{lem}[thm]{Lemma}{\bfseries }{\itshape }
\spnewtheorem{cor}[thm]{Corollary}{\bfseries }{\itshape }
\spnewtheorem{clm}{Claim}{\bfseries }{\itshape }
\numberwithin{clm}{thm}
\spnewtheorem{defn}{Definition}{\bfseries }{\itshape }

\newcommand{\Fig}[1]{#1.pdf}

\newcommand{\ShowTODO}[1]{{#1}}
\renewcommand{\ShowTODO}[1]{}

\newcommand{\TODOB}[2]{\ShowTODO{\todo[linecolor=#1,inline, backgroundcolor=#1!60!white,bordercolor=#1]{\sf #2}}}

\newcommand{\TODOJan}[1]{\TODOB{blue!60!white}{{\bfseries TODO Jan:} #1}}
\newcommand{\TODOYann}[1]{\TODOB{purple!70!white}{{\bfseries TODO Yann:} #1}}

\newcommand{\hle}[1]{{\begin{mdframed}[backgroundcolor=yellow,linewidth=0pt]
#1
\end{mdframed}}
}
\renewcommand{\hle}[1]{{#1}}
\newcommand{\hl}[1]{{#1}}

\newenvironment{dproof}{\begin{proof}}{\qed\end{proof}}

\newcommand{\E}[3]{E_{#1}(#2,#3)}
\newcommand{\Model}{\mathcal{M}}
\newcommand{\CompSec}[1]{\mathcal{S}_{#1}}
\newcommand{\Def}[1]{{\sf#1}}
\newcommand{\Op}{\text{{\tt(}}}
\newcommand{\Cp}{\text{{\tt)}}}
\newcommand{\ub}{\text{{\tt.}}}
\newcommand{\Ab}{{\sf A}}
\newcommand{\Cb}{{\sf C}}
\newcommand{\Gb}{{\sf G}}
\newcommand{\Ub}{{\sf U}}
\newcommand{\wP}{u}
\newcommand{\wS}{v}
\newcommand{\Sp}{S_{u}}
\newcommand{\Ss}{S_{v}}

\newcommand{\True}{{\sf True}}
\newcommand{\False}{{\sf False}}

\DeclareMathOperator*{\argmin}{argmin}

\newcommand{\Result}[1]{{\bf R#1}}

\newcommand{\MotifA}{{m}_{5}}
\newcommand{\MotifB}{{m}_{3\,\circ}}

\newcommand{\RNAFold}{\ensuremath{\text{\sf RNA-FOLD}}}
\newcommand{\UniqueFold}{\ensuremath{\text{\sf UNIQUE-FOLD}}}
\newcommand{\RNADesign}{\ensuremath{\text{\sf RNA-DESIGN}}}

\newcommand{\Designable}[1]{\text{\sf Designable}(#1)}
\newcommand{\Paired }[1]{\text{\sf Paired}(#1)}
\newcommand{\Unpaired }[1]{U_{#1}}
\newcommand{\MaxDeg}[1]{D({#1})}
\newcommand{\Sep}{{\sf Sep}}

\newcommand{\LevelFun}{L}
\newcommand{\Level}[1]{\LevelFun(#1)}

\newcommand{\Prefix}[2]{#2_{[1,#1]}}

\newcommand{\Substring}[3]{#3_{[#1,#2]}}

\def\mySecStr#1{\expandafter {\tt #1}\& }
\def\mySecStrAll#1{\ifx#1\mySecStrAll\else\mySecStr#1\expandafter\mySecStrAll\fi}
\def\mySeq#1{\expandafter {\relsize{-2}\sf #1}\&}
\def\mySeqAll#1{\ifx#1\mySeqAll\else\mySeq#1\expandafter\mySeqAll\fi}

\newcommand{\Problem}[3]{\begin{mdframed}[backgroundcolor=lightgray!10,
  linecolor=black,
  roundcorner=10pt,
  linewidth=0pt]
\noindent#1 problem\\
{\sf Input:} #2\\
{\sf Output:} #3
\end{mdframed} 
}
\newcommand{\WM}{\ensuremath{{\cal W}}}
\newcommand{\WMw}{\ensuremath{{\cal W}^{\mathrm{\kern0.05em w}}}}
\newcommand{\WMe}{\ensuremath{{\cal N}}}

\newcommand{\Deltae}{\ensuremath{\Delta_{\mathcal{N}}}}
\newcommand{\nGC }[2]{\ensuremath{n_{\Gb\Cb }(#1,#2)}}
\newcommand{\nGU }[2]{\ensuremath{n_{\Gb\Ub }(#1,#2)}}
\newcommand{\nAU }[2]{\ensuremath{n_{\Ab\Ub }(#1,#2)}}

\newcommand{\RNA}[4][]{
\begin{tikzpicture}[baseline={([yshift=-.5ex]rna)}]
  \matrix[matrix of nodes,nodes=cell,ampersand replacement=\&] (rna){
		\mySecStrAll #2 \mySecStrAll\\
		\ifthenelse{\equal{#3}{}}{}{ \mySeqAll #3 \mySeqAll\\}
		};
\ifthenelse{\equal{#4}{}}{}{%
	\foreach \x/\y in {#4}{\draw (rna-1-\x) edge[bp] (rna-1-\y);}%
}
\ifthenelse{\equal{#1}{}}{}{%
	\foreach \x/\y in {#1}{\draw (rna-2-\x) edge[altbp] (rna-2-\y);}%
}
\end{tikzpicture}}

\newcommand{\JRNA}[6][]{%
\begin{tikzpicture}[baseline={([yshift=-.5ex]rna)}]
  \matrix[matrix of nodes,nodes=cell,ampersand replacement=\&] (rna){
		\ifthenelse{\equal{#2}{}}{}{ \mySeqAll #2 \mySeqAll\\}
		};
\ifthenelse{\equal{#3}{}}{}{%
	\foreach \x/\y in {#3}{\draw (rna-1-\x) edge[jbp0] (rna-1-\y);}%
}
\ifthenelse{\equal{#4}{}}{}{%
	\foreach \x/\y in {#4}{\draw (rna-1-\x) edge[jbp1] (rna-1-\y);}%
}
\ifthenelse{\equal{#5}{}}{}{%
	\foreach \x/\y in {#5}{\draw (rna-1-\x) edge[jbp2] (rna-1-\y);}%
}
\ifthenelse{\equal{#6}{}}{}{%
	\foreach \x/\y in {#6}{\draw (rna-1-\x) edge[jbp3] (rna-1-\y);}%
}
\ifthenelse{\equal{#1}{}}{}{%
	\foreach \x/\y/\b/\a in {#1}{%
		\coordinate (X) at ($ (rna-1-\x)!.5!(rna-1-\y) $);%
		\draw ($ (X) - (0,\b) $) -- ($ (X) + (0,\a) $);%
	}%
}
\end{tikzpicture}}

 \journalname{Algorithmica}
\begin{document}

\title{Combinatorial RNA Design
}
\subtitle{Designability and Structure-Approximating Algorithm in Watson-Crick and Nussinov-Jacobson Energy Models}

\titlerunning{Combinatorial RNA Design}        

\author{Jozef Hale\v{s}\and Alice H\'eliou \and J\'an Ma\v{n}uch\and Yann Ponty \and Ladislav Stacho}

\authorrunning{Hale\v{s}, H\'eliou, Ma\v{n}uch, Ponty, and Stacho} 

\institute{
Yann Ponty$^{\ast}$ and Alice Héliou\at LIX (CNRS UMR 7161) Ecole Polytechnique \& Inria Saclay, Palaiseau, France\\
${\ast}$ Correponding author\\
              \email{yann.ponty@lix.polytechnique}
\and
Jozef Hale\v{s} and Ladislav Stacho\at Department of Mathematics, Simon Fraser University, Canada
\and 
J\'an Ma\v{n}uch\at Department of Mathematics, Simon Fraser University, Canada\\
             \emph{Present address:} Department of Computer Science, University of British Columbia, Canada
}

\date{Received: date / Accepted: date}

\maketitle

\begin{abstract} 
We consider the {\em Combinatorial RNA Design problem}, a minimal instance of RNA design where one must produce an RNA sequence that adopts a given secondary structure as its minimal free-energy structure. We consider two free-energy models where the contributions of base pairs are additive and independent: the purely combinatorial {\em Watson-Crick model}, which only allows equally-contributing $\Ab-\Ub$ and $\Cb-\Gb$ base pairs, and the real-valued {\em Nussinov-Jacobson model}, which associates arbitrary energies to $\Ab-\Ub$, $\Cb-\Gb$ and $\Gb-\Ub$ base pairs.

We first provide a complete characterization of designable structures using restricted alphabets and, in the four-letter alphabet, provide a complete characterization for designable structures without unpaired bases. When unpaired bases are allowed, we characterize extensive classes of (non-)designable structures, and prove the closure of the set of designable structures under the stutter operation. Membership of a given structure to any of the classes can be tested in $\Theta(n)$ time, including the generation of a solution sequence for positive instances.

Finally, we consider a structure-approximating relaxation of the design, and provide a $\Theta(n)$ algorithm which, given a structure $S$ that avoids two trivially non-designable motifs, transforms $S$ into a designable structure constructively by adding at most one base-pair to each of its stems. 

\end{abstract}

\begin{acknowledgements}
The authors thank Cédric Chauve for fruitful discussions and constructive criticisms.
YP is supported by the Pacific Institute for the Mathematical Sciences (PIMS), and the French Agence Nationale de la Recherche (ANR-14-CE34-0011).
\end{acknowledgements}


\section{Introduction}
\label{sec:introduction}
RiboNucleic Acids (RNAs) are biomolecules which act in almost every aspect of cellular life, and can be abstracted as a sequence of nucleotides, i.e., a string over the alphabet $\{\Ab,\Ub,\Cb,\Gb\}$.
Due to their versatility, and the specificity of their interactions, they are increasingly being used as therapeutic agents~\cite{Wu2014}, and as building blocks for the emerging field of synthetic biology~\cite{Rodrigo2013,Takahashi2013}. A substantial proportion of the functional roles played by RNA rely on interactions with other molecules to activate/repress dynamical properties of some biological system, and ultimately require the adoption of a specific conformation. Accordingly, RNA bioinformatics has dedicated much effort to developing energy models~\cite{Mathews1999,Turner2010} and algorithms~\cite{Nussinov1980,Zuker1981} to predict the \Def{secondary structure of RNA}, a combinatorial description of the conformation adopted by an RNA which only retains interacting positions, or base pairs.
Historically, structure prediction has been addressed as an optimization problem, whose expected output is a secondary structure which minimizes some notion of free-energy~\cite{Nussinov1980,Zuker1981}. The performances of the RNA folding prediction problem have now reached a point where {\em in silico} predictions are generally considered reliable~\cite{Mathews1999}, allowing for large scale studies and fueling the discovery of an increasing number of functional families~\cite{Griffiths-Jones2003}.

Due to the existence of expressive, yet tractable, energy models and conformational spaces, coupled with promising applications in multiple fields (pharmaceutical research, natural computing, biochemistry\ldots), a wide array of computational methods~\cite{Hofacker1994,Busch2006,Aguirre-Hernandez2007,Dai2009,Avihoo2011,Taneda2011,Zadeh2011,Levin2012,Lyngso2012,Garcia-Martin2013a,HoenerZuSiederdissen2013,Reinharz2013,Zhou2013,Bau2013,Esmaili-Taheri2014}
 has been proposed to tackle the natural inverse version of the structure prediction, the \Def{RNA design problem}. In this problem, one attempts to perform the {\em in silico} synthesis of artificial RNA sequences,  performing a predefined biological function {\em in vitro} or {\em in vivo}. Given the prevalence of structure in the function of an RNA, one of the foremost goal of RNA design (sometimes named  \Def{inverse folding} in the literature) is to ensure that the designed sequence folds into a predefined secondary structure, preferentially to any alternative structure. In other words, the chosen conformation should not be challenged by alternative stable structures having similar or lower free-energy. 

Despite a rich, fast-growing, body of literature dedicated to the problem, there is currently no exact polynomial-time algorithm for the problem. Moreover, the complexity of the problem remains open (see Section~\ref{sec:conclusion} for a discussion). It can be argued that this situation, quite exceptional in the field of computational biology, partly stems from the intricacies of the Turner free-energy model~\cite{Turner2010} which associates experimentally-determined energy contributions to $\sim2.4\times 10^4$ structure/sequence motifs. This motivates a reductionist approach, where one studies an idealized version of the RNA design problem, lending itself to algorithmic intuitions, while hopefully retaining the presumed difficulty of the original problem and provides intuitions for future studies of the problem under more sophisticated energy models.

In this work, we introduce the {\em Combinatorial RNA Design problem}, a {\em minimal} instance of the RNA design problem which aims at finding a sequence that admits the target structure as its unique base pair maximizing structure. 
After this short introduction, Section~\ref{sec:defs} states definitions and problems. In Section~\ref{sec:results}, we state our main results and prove them in Section~\ref{sec:proofs}, including an extended weighted version that allows additional types of base pairs. Finally, we conclude in Section~\ref{sec:conclusion} with some remarks, open problems and future extensions of this work.

\section{Definitions and notations}\label{sec:defs}
\begin{figure}
{\centering
\includegraphics[width=\linewidth]{\Fig{fig1}}\\}

\caption{Four equivalent representations for an RNA secondary structure of length $68$, consisting of $20$ base pairs forming $7$ bands: outer-planar graph ({\sf \bfseries a.}), arc-annotated representation ({\sf \bfseries b.}), parenthesized expression ({\sf \bfseries c.}), and tree representation ({\sf \bfseries d})\TODOYann{Replace with Jano's simpler example}\label{fig:reps}}
\end{figure}

\paragraph{RNA secondary structure.}
An RNA can be encoded as a \Def{sequence} of nucleotides, i.e., a string $w = w_{1}\cdots w_{|w|}\in\{\Ab,\Ub,\Cb,\Gb\}^\star$. The \Def{prefix of $w$ of length $i$} is denoted as $\Prefix{i}{w}$ and $|w|_{b}$ denotes the number of occurrences of $b$ in $w$.
A \Def{(pseudoknot-free) secondary structure} $S$ on an RNA of length $n$ is a pair $(n,P)$, where $P$ is a set of \Def{base pairs} $\{(l_{i},r_{i})\}_{i=1}^p \subset [1,n]^2$ such that:
\begin{itemize}
\item $\forall i\in [1,p]$, $l_i<r_i$;
\item Each position is involved in at most
  one base pair, {\it i.e.} $\forall i\ne j\in [1,p], l_{i}\ne l_{j}$, $l_{i}\ne
  r_{j}$, $r_{i}\ne r_{j}$;
\item Base pairs are pairwise \Def{non-crossing},{\it i.e.} $\nexists i,j\in [1,p]$, $l_{i} < l_{j} < r_{i} < r_{j}$,  .
\end{itemize}
The set of \Def{unpaired positions} $\Unpaired{S}$ of a secondary structure $S = (n,P)$ is the set of indices $k\in[1,n]$ that are not involved in any of the base pairs in $S$. A structure $S$ is called \Def{saturated} if and only if its positions are all paired, {\it i.e.} iff $\Unpaired{S} = \emptyset $. Conversely, a structure $S$ is \Def{empty} when none of its positions is paired,  {\it i.e.} iff $S = (n,\emptyset )$.

The \Def{set of all secondary structures} is denoted by $\CompSec{}$, and its restriction to structures of length $n$ by $\CompSec{n}$. 
Secondary structures are typically expressed using a variety of equivalent representations, illustrated by Figure~\ref{fig:reps} and formally defined further in this section.

\TODOYann{Review next notions}
Given a sequence $w$ and a structure $S = (|w|,P)$, let $u_{i} = \varepsilon $ if $i\in \Unpaired{S}$ and $u_{i} = w_{i}$, otherwise, where $\varepsilon $ is the empty sequence. 
\hle{Define the \Def{$S$-paired restriction} of $w$, denoted by $\Paired{w,S}$, as the subsequence of $w$ consisting of the paired positions only, \textit{i.e.}  $\Paired{w,S}=u_{1}\cdots u_{|w|}$.
Similarly, define the \Def{paired restriction of $S = (n,P)$}, denoted by  $\Paired{S}$, as the substructure of $S$ consisting of paired positions only, \textit{i.e.} $\Paired{S}=(|\Paired{w,S}|,\{(|u_{1}\cdots u_{i}|,|u_{1}\cdots u_{j}|)\mid (i,j)\in P\})$, where $w$ is any sequence of length $n$ and $u_{i}$'s are defined as above.}

A maximal subset $B = \{(i,j),(i + 1,j - 1),\dots,(i + \ell - 1, j - \ell + 1)\}$ of $P$ for some integer $i,j,\ell $ is called a \Def{band} (also referred to as \Def{helix} or \Def{stem} in related works) of  \Def{size} $\ell = |B|$, of $S = (n,P)$. Note that every base pair belongs to exactly one band. In other words, the base-pairs of a secondary structure can be unambiguously partitioned into a set of bands. 

\paragraph{Dot-parentheses notation.}
A \Def{well-parenthesized sequence} $s\in \{\Op,\Cp,\ub\}^{*}$ can be used to represent a secondary structure. There exists a one-to-one correspondence between secondary structures and such well-parenthesized sequences: any base pair $(l,r)\in P$ becomes a pair of corresponding opening and closing parentheses in $s$ at position $l$ and $r$ respectively ($s_l = \Op$ and  $s_r = \Cp$), and any unpaired position $i$ corresponds to a dot ($s_i=\ub$). This representation is illustrated by Figure~\ref{fig:reps}.c. A \Def{concatenation} of two structures $S$ and $S'$, denoted by $S.S'$ or simply $SS'$ wherever unambiguous, is the structure corresponding to the well-parenthesized sequence obtained by concatenating the well-parenthesized sequences of $S$ and $S'$.

\paragraph{$k$-stutter.}
The $k$-\Def{stutter} of a sequence $s$, denoted by $s^{[k]}$ is the result of an independent copy $k$-times of each of the characters in $s$. For instance, the 3-stutter of a sequence $\Ab\Ub\Ub\Cb$ is $\Ab\Ab\Ab\Ub\Ub\Ub\Ub\Ub\Ub\Cb\Cb\Cb$. This operation also applies to an RNA structure $S$, and $S^{[k]}$ denotes the RNA structure obtained by applying the usual $k$-stutter to the dot-parentheses representation of $S$.

\paragraph{Tree representation.}
Alternatively, the \Def{tree representation}, denoted by $T_{S}$, for $S = (n,P)$ is a rooted ordered tree whose vertex set $V_S$ consists of intervals $[l,r]$ for any base pair $(l,r)\in P$, and $[k,k]$ for every $k\in \Unpaired{S}$.
A \Def{virtual root} $[0,n+1]$ is added for convenience. Any node labeled by $[k,k]$ is called \Def{unpaired}, and any other node (including the virtual root) are considered as \Def{paired}. The \Def{children} of an interval $I \in V_S$ are the maximal proper subintervals $I'\in V_S$ of $I$ ordered by the left points of the intervals. 
The \Def{degree} of a vertex $I\in V_S$ is the total number of its paired neighbors, including its parent (if any).
We denote by $\MaxDeg{S}$ the maximal degree of nodes in $T_{S}$. Figure~\ref{fig:reps}.d shows the tree representation of a typical secondary structure. 

\paragraph{Proper, greedy and separated coloring of the tree representation.}
Consider the tree representation $T_{S}$ of structure $S$. A \Def{coloring} of $T_{S}$ associates a color, chosen among black, white, or gray, to each paired node of $T_{S}$ that is different from the root. This coloring is called \Def{proper} if:
\begin{enumerate}
\item[i)] Each node has at most one black child, at most one white child, and at most two grey children;
\item[ii)] Any $c$-colored node has at most one $c$-colored child;
\item[iii)] Black nodes shall not have a white child, and white nodes shall not have a black child.
\end{enumerate}
A \Def{greedy coloring} of $T_{S}$ is the coloring obtained by recursive application of the following rule starting from the root and continuing towards leaves: if the node is black, color the first paired child black and the remaining paired children gray, if the node is white, color the first paired child white and the remaining paired children gray, otherwise (the gray node or the root), color the first paired child black, second white and the remaining paired children gray. It is easy to check that if the degree of each node is at most four then the greedy coloring is a proper coloring.

Given a proper coloring of $T_{S}$, let the \Def{level} of each node be the number of black nodes minus the number of white nodes on the path from this node to the root. A proper coloring is called \Def{separated} if the two sets of levels, associated with gray and unpaired nodes respectively, do not overlap.

\subsection{Statement of the generic RNA design problem}

Consider an \Def{energy model} $\Model$, which associates a \Def{free-energy} $\E{\Model}{w}{S}\in \mathbb{R}^-\cup \{+\infty\}$ to each secondary structure $S\in \CompSec{|w|}$ for a given RNA sequence $w$.
The \Def{minimum free-energy (MFE) structure prediction} problem is typically defined as follows:
\Problem{\RNAFold$_{\Model}$}%
{RNA sequence $w$}%
{$S_{\Model }^\star (w) := \argmin_{S'\in \CompSec {|w|}} \E{\Model}{w}{S'}\,,$}
where $\argmin $ returns a single structure $S_{\Model }^\star (w)$, arbitrarily chosen amonst those having minimum free-energy. 

The existence of alternative competing structures, {\it i.e.} one or several secondary structure(s) having (almost) minimal free-energy for a given RNA, impacts the efficacy of the folding process. The detection of such situations is therefore of interest, and can be rephrased as the following problem:
\Problem{\UniqueFold$_{\Model}$}%
{Sequence $w$ + Energy distance $\Delta>0$}%
{\True{} if, for every
$S'\in \CompSec{|w|}\setminus\{S_\Model^\star(w)\}$, one has:
$$\E{\Model}{w}{S'} \ge \E{\Model}{w}{S_\Model^\star(w)}+\Delta \,.$$
\False{} otherwise.}

We can now define the \Def{combinatorial RNA Design problem} as:
\Problem{\RNADesign$_{\Model,\Sigma}$}%
{Secondary structure $S$ + Energy distance $\Delta>0$}%
{RNA sequence $w\in\Sigma^\star$ -- called an \Def{$(\Model ,\Sigma ,\Delta )$-design for $S$} -- such that:\\
$$ \RNAFold_\Model(w)=S\ \text{ and }\ \UniqueFold_\Model(w,\Delta),$$
or {$\varnothing$} if no such sequence exists.}

The structures for which there exists an $(\Model ,\Sigma ,\Delta )$-design are called \Def{$(\Model ,\Sigma ,\Delta )$-designable}. Let $\Designable{\Model ,\Sigma ,\Delta } $ be the set of all such structures. If it is clear from the context, we will usually drop $\Model $, $\Sigma $ and/or $\Delta $.

\begin{figure}
  {\centering
  \includegraphics[width=\linewidth]{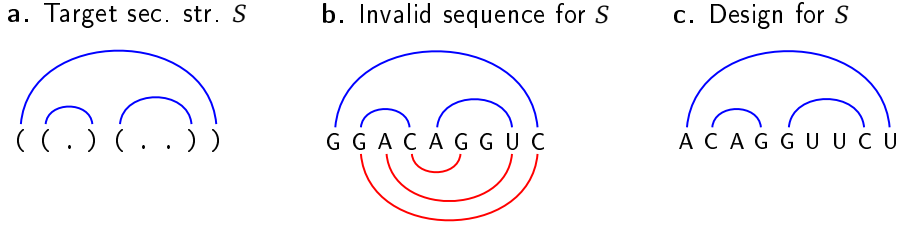}\\}
  
  \caption{The combinatorial RNA design problem: Starting from a secondary structure $S$ ({\sf \bfseries a.}), our goal is to design an RNA sequence which uniquely folds, with maximum number of base pairs, into $S$. The sequence proposed in {\sf \bfseries b.} is invalid due to the existence of an alternative structure (lower half-plane, red) having the same number of base pairs as $S$. The right-most sequence ({\sf \bfseries c.}) is a design for $S$.}
\end{figure}

\subsection{Base pair sum energy models}

In this work, we will consider two types of \Def{base pair sum} energy models, where the free-energy of a structure is simply obtained by sum, over all base pairs, individual independent contributions associated with each pair. 
%

\begin{defn}[Base pair sum energy model $\mathcal{M}$]
Let $w$ be an RNA sequence and $S$ a secondary structure in $\CompSec{|w|}$. Then
$$\E{\mathcal{M}}{w}{S} = \sum_{(l,r)\in S} E_{\mathcal{M}}(w_{l},w_{r})\,,$$
where $E_{\mathcal{M}}(x,y)$ is the energy induced by a base pair $x - y$.
\end{defn}

To define a model of interest, it is sufficient to specify the energies of base pairs.
First, we consider a minimal energy model, named \Def{Watson-Crick model} due to its similarity with the DNA base-pairing rules.
The model is purely combinatorial, as it associates a homogenous $-1$ energy contribution to each valid base-pair, and only allows $\Gb-\Cb$ and $\Ab-\Ub$ to pair.

\begin{defn}[Watson-Crick energy model $\WM$]
\[E_{\WM }(x,y) =
  \begin{cases}
    -1 &\text{if $\{x,y\}=\{\Gb,\Cb\}$ or $\{x,y\}=\{\Ab,\Ub\}$}\\
    +\infty&\text{otherwise.}
  \end{cases}
\]
\end{defn}


A more general model, named the \Def{Nussinov-Jacobson model}, allows $\Gb-\Ub$ base-pairs to occur, and associates arbitrary weights to the base pairs depending on their content. It is named after the authors of the first polynomial-time algorithm for predicting the MFE under a similar energy model~\cite{Nussinov1980}.

\begin{defn}[Nussinov-Jacobson energy model $\WMe$]
\[E_{\WMe }(x,y) =
  \begin{cases}
    \alpha &\text{if }\{x,y\}=\{\Gb,\Cb\}\\
    \beta &\text{if }\{x,y\}=\{\Ab,\Ub\}\\
    \gamma &\text{if }\{x,y\}=\{\Gb,\Ub\}\\
    +\infty&\text{otherwise.}
  \end{cases}
\]
where $\alpha ,\beta,\gamma  \in  \mathbb{R}^-$ and $\alpha,\beta<\gamma$.
\end{defn}

Note that the last condition of the above definition is typically satisfied by empirical estimates of base pair energies. Namely, $\Gb - \Ub$ base pairs, also named \Def{Wobble base pairs}, are much weaker than its alternatives. They are mediated by a single hydrogen bond, as opposed to $2$ (resp. $3$) bonds for $\Ab-\Ub$ (resp. $\Gb-\Cb$).

We say that the structure is \Def{compatible} with a sequence $w$, according respect to an energy model ${\cal M}$, if $\E{\mathcal{M}}{w}{S} < +\infty $.

It is worth noting that minimizing $\E{\WM}{w}{S}$ is equivalent to maximizing $|S|$, thus \RNAFold$_{\WM}$ is a classic base pair maximization problem. Moreover, both \RNAFold$_{\WM}$ and \RNAFold$_{\WMe}$ can be solved in polynomial time using dynamic programming, historically in $\mathcal{O}(n^3)$ complexity ~\cite{Nussinov1980}, or in $\mathcal{O}(n^3/\log(n))$ current best time complexity~\cite{Frid2010}.
A backtracking procedure reconstructs the structure having minimal energy, and can be easily
adapted to provide a $\Theta(n^3)$ algorithm for the \UniqueFold$_{\Model}$ problem.

\section{Statement of the results}\label{sec:results}

In this section, we characterize sets of secondary structures which can or cannot be designed using (a subset of) $\{\Ab , \Cb,\Gb,\Ub \}$, a (restricted) set of base pairs and a desired energy distance $\Delta$ within an energy model $\cal M$. The proofs of our statements are largely interconnected, and have been regrouped in Section~\ref{sec:proofs}.

First, let us remark that the empty secondary structures are the only ones that are designable for arbitrary large energy distances $\Delta$.

\begin{thm}\label{th:limit}
  For any ${\cal M}\in \{{\cal W},{\cal N}\}$, and any energy distance $\Delta$ such that \begin{equation}\Delta  > -E_{\cal M}({\sf X},{\sf Y}),\quad \forall {\sf X},{\sf Y}\in \{\Ab , \Cb,\Gb,\Ub \}^2,\label{eq:condDelta}\end{equation}
only the empty secondary structures are designable.
\end{thm}

\begin{dproof}
For any non-empty secondary structure $S$ having energy $E$ on some sequence $w$, removing any base pair ${\sf X}-{\sf Y}$ from $S$ yields an alternative structure $S'$ whose energy is $E'= E-E_{\cal M}({\sf X},{\sf Y}) < E + \Delta $. In other words, $S'$ is a competing structure  at distance less than $\Delta$ of $S$, {\it i.e.} $w$ is not a valid $\Delta$-design for $S$. 

Moreover, any empty structure of length $n$ is designable. Indeed, none of the models allows for pairs of the form ${\sf X}-{\sf X}, {\sf X}\in\{\Ab ,\Cb,\Gb,\Ub \}$, so any sequence ${\sf X}^n$ admits the empty structure, having $0$ energy, as its only secondary structure having finite free-energy, {\it i.e.} ${\sf X}^n$ is a design for the empty structure for any finite $\Delta > 0$.
\end{dproof}

%
%

\subsection{Watson-Crick model ${\cal W}$ ($\Delta=1$)}
We provide (partial) characterizations for the sets $\Designable{\Sigma}$ of designable structures over partial alphabets $\Sigma$ in the $\cal W$ model. From Theorem~\ref{th:limit}, combined with the purely combinatorial nature of the energy model, we observe that non-trivial structures are designable only when $\Delta\in (0,1]$, and that the set of designable structures is unaffected by the precise value of $\Delta$ on the segment. Therefore we set $\Delta=1$ without loss of generality, and consider the structures whose designed sequences lose at least one base-pair when forming alternative structures. We obtain the following meta-theorem.
\begin{thm}
  In the Watson-Crick energy model $\cal W$, and assuming an energy distance of $\Delta=1$, results \Result{1} through \Result{8} hold.
\end{thm}

Let $\Sigma_{c,u} $ be an alphabet with $c$ pairs of complementary bases and $u$ bases without a complementary base. Without loss of generality in the ${\cal W}$ model, we will assume that $\Sigma_{1,0} = \{\Gb ,\Cb \} $ and $\Sigma_{1,1} = \{\Gb,\Cb ,\Ab \}$.

\subsubsection{Designability over restricted alphabets.}

\begin{figure}
  {\centering
  \includegraphics[width=.7\textwidth]{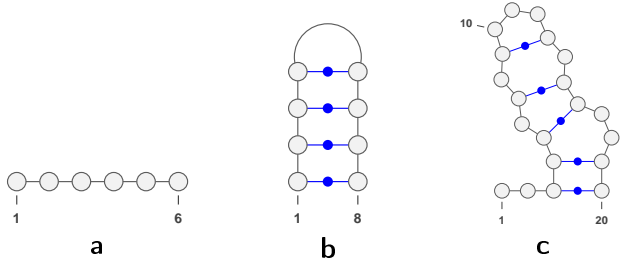}\\}
  
  \caption{Examples of structures satisfying conditions of \Result{1}  ({\bf a.}), \Result{2}  ({\bf b.}) and \Result{3}  ({\bf c.}).}
  \label{fig:examplesR1-3}
\end{figure}

First, we provide a complete characterization of designable (secondary) structures using an alphabet $\Sigma$ of restricted cardinality: 
\begin{enumerate}
\item[\Result{1}] 
  For every $u\in\mathbb{N}^{+}$, $\Designable{\Sigma_{0,u} } = \{(n,\emptyset)\mid\forall n\in\mathbb{N}\}$;
\item[\Result{2}] 
  $\Designable{\Sigma_{1,0}} = \{S\in\CompSec{} \mid S \text{ is saturated and } \MaxDeg{S}\le 2\} \cup \{(n,\emptyset)\mid\forall n\in\mathbb{N}\}$;
\item[\Result{3}] 
  $\Designable{\Sigma_{1,1}} = \{S\in\CompSec{} \mid \MaxDeg{S}\le 2\}$.
\end{enumerate}

Figure~\ref{fig:examplesR1-3} shows examples of secondary structures satisfying conditions of these three results.

\subsubsection{Designability over the complete alphabet $\Sigma_{2,0} =  \{\Ab,\Ub,\Cb,\Gb\}$.}

\begin{figure}
  {\centering
  \includegraphics[width=.7\textwidth]{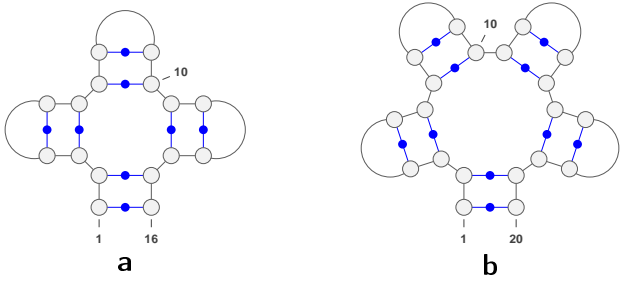}\\}

  \caption{Examples of two saturated structures, one satisfying conditions of \Result{4} (\textsf{\textbf{a.}}) and one that does not (\textsf{\textbf{b.}}). The structure on the left has pair degree 4 and is designable, and the structure on the right has pair degree 5 and is not designable.}
  \label{fig:examplesR4}
\end{figure}

We first characterize the set of designable saturated structures, {\it i.e.} structures whose positions are all paired to some other position.
\begin{enumerate}
\item[\Result{4}] 
  $\{S\in\Designable{\Sigma_{2,0}} \mid S \text{ is saturated} \} = \{S\in\CompSec{} \mid \MaxDeg{S}\le 4 \text{ and }  S \text{ is saturated} \}$.
\end{enumerate}

Figure~\ref{fig:examplesR4} shows two saturated secondary structures one with pair degree 4 and one with pair degree 5. By \Result{4}, the former is $\Sigma_{2,0}$-designable, while latter is not.

\begin{figure}
  {\centering
  \includegraphics [width=.7\textwidth]{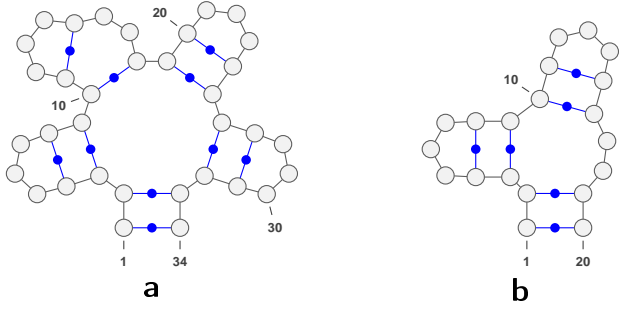}\\}
  
  \caption{Examples of two structures containing motifs $\MotifA $ ({\bf a.}) and $\MotifB$ ({\bf b.}), respectively. By \Result{5} these two structures are not designable.}
  \label{fig:examplesR5}
\end{figure}

When unpaired positions are allowed in the target structure, our characterization is only partial:
\begin{enumerate}
\item[\Result{5}] 
  Let $\MotifA$ represent ``a node having degree more than four'', and $\MotifB$ be ``a node having one or more unpaired children, and degree greater than two'' (cf. Figure~\ref{fig:examplesR5}), then 
$$\Designable{\Sigma_{2,0}} \cap  \{S\in\CompSec{} \mid \text{$S$ contains $\MotifA$ or $\MotifB$}\} = \emptyset \,;$$
\item[\Result{6}] 
  Let $\Sep$ be the set of structures for which there exists a separated (proper) coloring of the tree representation, then $\Sep\subset \Designable{\Sigma_{2,0}}$;
\item[\Result{7}] 
  The set of $\Sigma_{2,0}$-designable structures is closed under the $k$-stutter operations:
$$ \forall S\in \CompSec{}, \forall k\in \mathbb{N}^+:\quad S\in \Designable{\Sigma_{2,0}} \implies S^{[k]} \in \Designable{\Sigma_{2,0}}\,.$$
\end{enumerate}

\begin{figure}
\centering 
  \includegraphics[width=\textwidth]{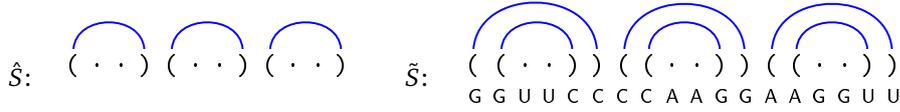}
  \caption{An example of non-designable (left) and designable structure (right).}
  \label{fig:struct:input}
\end{figure}

We note however that reverse implication is not true: $S^{[k]}\in \Designable{\Sigma_{2,0}}$ does not imply that $S\in \Designable{\Sigma_{2,0}}$. For instance, it is easily verified that $\hat S^{[2]}$ is $\Sigma_{2,0}$-designable, while $\hat S$ is not, as shown in Figure~\ref{fig:struct:input}. Membership to the classes described in \Result{1}-\Result{5} can be tested by trivial linear-time algorithms. 
These algorithms can also be easily adapted into linear-time algorithms for the production of a concrete design, thereby offering partial solutions to the  \RNADesign$_{\Model,\Sigma}$ problem.

\subsubsection{Structure-approximating algorithm.} 
\begin{figure}
{\centering \includegraphics[width=.95\textwidth]{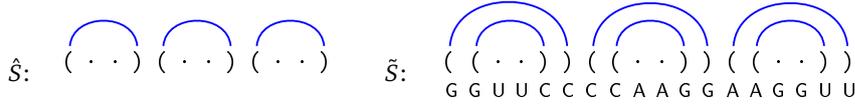}\\}

\caption{Application of the structure-approximating algorithm to the non-designable structure $\hat S$ in Fig.~\ref{fig:struct:input}:
A base pair (circled black node) is inserted in the greedily colored tree, offsetting the levels of white and unpaired nodes (crosses) to even and odd levels respectively, so that the resulting tree is proper/separated, representing a designable structure.}
\label{fig:inflation}
\end{figure}

Unfortunately, avoiding $\MotifA$ and $\MotifB$, while necessary, is generally not sufficient to ensure designability. For instance, consider $\hat S$ in Figure~\ref{fig:struct:input}
clearly does not contain $\MotifA$ or $\MotifB$, yet it cannot be designed.
In such cases, unwanted interactions can be somehow penalized by duplicating some base pairs. For instance, duplicating a single base pair in $\hat S$ yields a $\Sigma_{2,0}$-designable structure $\tilde S$, as shown by Figure~\ref{fig:inflation}.
\begin{enumerate}
\item[\Result{8}] Any structure $S$ avoiding $\MotifA$ and $\MotifB$ can be transformed in $\Theta(n)$ time into a $\Sigma_{2,0}$-designable structure $S'$. This is done by duplicating a subset of the base pairs of $S$, at most one per band, such that the greedy coloring of the resulting structure is proper and separated, as illustrated by  Figure~\ref{fig:inflation}.
\end{enumerate}
%
%
%
%
%
%
%

\subsection{Nussinov-Jacobson energy model $\cal N$ ($\Delta\le \min (|\alpha |,|\beta |)]$)}\label{sec:nj}

We consider the validity of the above results in the Nussinov-Jacobson model. Note that the consideration of $\Gb-\Ub$ base pairs, by loosening the notion of complementarity, forces us to abuse our notation for $\Sigma_{i,j}$. Namely, $\Sigma_{2,0}$ is taken to represent the full alphabet, even though it now strictly allows three types of base pairs.
\begin{thm} For any $\Delta\in (0,\min (|\alpha |,|\beta |)]$, statements \Result{1} through 
\Result{8} hold in the Nussinov-Jacobson model.
\end{thm}

\section{Proofs}\label{sec:proofs}

\subsection{Watson-Crick model ($\Delta=1$)}
\Result{1} is trivial since, in the absence of complementary letters, empty structures are the only one whose energy is not infinite. 
%
\begin{thm}[$\Rightarrow$ Result \Result{2}, \Result{3} and \Result{4}]
  \label{t:result4}
  A saturated secondary structure $S$ is $\Sigma_{c,0}$-designable if and only if $\MaxDeg{S}\le 2c$.
\end{thm}

\begin{dproof}
First, we will show that the degree condition is necessary. Assume to the contrary that $\MaxDeg{S} > 2c$ and $S$ has a design $w$. Let $[a,b]$ be a vertex with degree $d\ge 2c + 1$ in $T_{S}$. Let $\{[l_{i},r_{i}]\}_{i = 1 }^{d}$ be the (paired) children of $[a,b]$ and the node $[a,b]$ if $[a,b]$ is not the root. Let $L_{i} = l_{i}$ and $R_{i} = r_{i}$ if $[l_{i},r_{i}]$ is a child of $[a,b]$, and $L_{i} = r_{i}$ and $R_{i} = l_{i}$ if it is $[a,b]$. Then among bases $w_{L_{1}},\dots,w_{L_{d}}$ must be a pair of repeated letters. Let $w_{L_{i}} = w_{L_{j}}$ be such a pair with $L_{i} < L_{j}$. It is easy to check that $S\setminus \{(l_{i},r_{i}),(l_{j},r_{j})\} \cup \{(L_{i},R_{j}),(R_{i},L_{j})\}$ is a structure compatible with $w$ with the same number of base pairs as $S$, a contradiction with the assumption that $w$ is a design for $S$.

To show that the degree condition is also sufficient, we need further definitions and claims.
First, we say that a sequence $w\in\Sigma^{*}$ is \Def{saturable} if there is a saturated structure compatible with $w$. Note that the concatenation of two saturable sequences is also saturable. Then the following claim characterizes the cases when a saturable sequence can be split into saturable sequences.

\begin{clm}\label{l:saturated}
Let $w = \wP \wS$ be a saturable sequence of length $k$. If $\wP$ is saturable, then so is $\wS$.
\end{clm}
\begin{dproof}
Consider a saturated structure $S$ compatible with sequence $w$ and a saturated structure $\Sp$ compatible with $\wP$. We will construct a saturated structure $\Ss$ compatible with $\wS$.

Consider a graph $G$ with vertex set $\{1,\dots,k\}$ and edge set defined by pairs in $S\cup \Sp$. Obviously, this graph is a collection of alternating paths (alternating between pairs from $S$ and from $\Sp$, starting and ending with positions in $\wS$) and alternating cyclic paths, and it has a planar embedding such that all vertices lie on a line in their order: pairs in $S$ are drawn as non-crossing arcs above the line and pairs in $\Sp$ as non-crossing arcs below the line. Note that every position in $\wS$ is an end-point of exactly one path in the collection.

Define set of base pairs $\Ss$ by pairing the end-points of the paths in $G$, cf. Figure~\ref{fig:saturated}. We will show that $\Ss$ is a structure. Consider a graph $G'$ constructed by adding pairs in $\Ss$ to $G$. This graph is a collection of cyclic paths. Consider an embedding of $G'$ into plane that extends the planar embedding of $G$ by adding arcs corresponding to the pairs in $\Ss$ below the line containing all the vertices. If two base pairs $b,b'\in \Ss$ cross then the cyclic path containing $b$ and the cyclic path containing $b'$ intersect in exactly one point. By Jordan's curve theorem, this is a contradiction. It follows that $\Ss$ is a saturated structure, and hence $\wS$ is also saturable.
\end{dproof}

We define $w$ to be an \Def{atomic saturable sequence} if no proper prefix
of $w$ is saturable. 
Every saturated structure compatible with an atomic saturable sequence $w$ contains the base pair $(1,|w|)$, \hl{since otherwise it contains the base pair $(1,j)$, with $j<|w|$, and consequently $w_{[1,j]}$ is a saturable proper prefix of $w$}. On the other hand, by Claim~\ref{l:saturated}, if every saturated structure compatible with $w$ contains the pair $(1,|w|)$, then $w$ is an atomic saturable sequence. A design $w$ that is also an atomic saturable sequence will be called an \Def{atomic saturable design}.
\begin{figure}
\centering
  {\centering \includegraphics{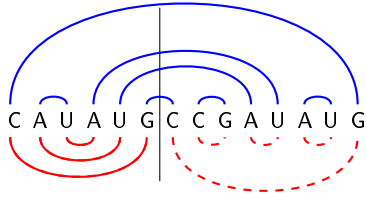}\\}
  
\caption{Construction of the saturated structure compatible with the suffix $\wS$. The vertical line splits the sequence into a prefix $\wP$ and a suffix $\wS$. Top (blue) and bottom (red) arcs depict saturated structures compatible with $w$ and $\wP$ respectively. Dashed arcs represent the induced saturated structure compatible with $\wS$, \hl{they are connecting end-points of the alternating  bottom/top full path}.}
\label{fig:saturated}
\end{figure}
A concatenation of two or more atomic saturable designs is obviously not an atomic saturable sequence and it is not necessarily a design. However, we have the following claims.
\begin{clm}\label{l:atomic1}
The concatenation of $t$ atomic saturable designs $w^{1},\ldots ,w^{t}$ for structures $S^1,\ldots ,S^{t}$, such that $w^{i}_{1}\ne w^{j}_{1}, \forall 1\le i < j\le t$, is a design for the concatenated (saturated) structure $S=S^1\cdots S^{t}$.
\end{clm}
\begin{dproof}
Assume that $W := w^{1}\cdots w^{t}$ is not a design, then there exist a saturated structure $S'\neq S$ for $W$. We show that positing such an alternative structure leads to a contradiction. Recall that each $S^{i}$ is saturated and contains a pair $(1, |w^{i}|)$. If $S'$ pairs the first and last letters in each $w^{i}$, $i\in[1,t]$, then $S' = S$ since each $w^{i}$ is a design, a contradiction.
Let $w_{i}$ be the leftmost sequence such that $w^{i}_1$ is not paired with $w^{i}_{|w^{i}|}$ in $S'$. Since $S'$ must be also saturated, $w^{i}_1$ must be paired. Let $w^{j}_{k}$, $j\ge i$, be the partner of $w^{i}_1$ in $S'$, and let $u := w^{i}\cdots w^{j-1}\Prefix{k}{w^{j}}$. If $k = |w^{j}|$, then $j > i$ and, by complementarity, $w^{i}_{1} = w^{j}_{1}$ which contradicts the preconditions. Hence, we can assume that $k < |w^{j}|$. Since $u$ and each of the $w^{i},\dots,w^{j - 1}$ are saturable, by iterated application of Claim~\ref{l:saturated}, we conclude that $v = \Prefix{k}{w^{j}}$ is saturable as well. 
This contradicts the precondition that $w^{j}$ is an atomic saturable design, since $v$ is a proper prefix of $w^{j}$.  
We conclude that no alternative saturated folding exists for $W$, i.e., $W$ is a design for $S$.
\end{dproof}
\begin{clm}\label{l:atomic2}
  Consider $t$ atomic saturable designs $w^{1}=w^{1}_1\cdots w^{1}_{|w^{1}|}$, \dots, $w^{t}=w^{t}_1\cdots w^{t}_{|w^{t}|}$ and a pair $a,b$ of complementary letters such that $w^{i}_{1}\ne b$ for every $1\le i\le t$ and $w^{i}_{1}\ne w^{j}_{1}$ for every $1\le i < j\le t$. Then $W = aw^{1}\cdots w^{t}b$ is an atomic saturable design.
\end{clm}
\begin{dproof}
We will first show that $W$ is an atomic saturable sequence. Assume to the contrary that there is a proper prefix of $W$ that is saturable. Consider the shortest such prefix $aw^{1}\cdots w^{i}\Prefix{j}{w^{i + 1}}$ with $1 \leq j < |w^{i+1}|$ and $1\leq i <t$. Obviously, $a$ has to be paired with $w_{j}^{i + 1}$, otherwise we can find a shorter saturable prefix. This implies that $b = w_{j}^{i + 1}$ and that $w^{1}\cdots w^{i}\Prefix{j - 1}{w^{i + 1}}$ is saturable as well. By repeated application of Claim~\ref{l:saturated}, we have that $\Prefix{j - 1}{w^{i + 1}}$ is saturable. Since it is a prefix of atomic saturable sequence $w^{i + 1}$, it must be the empty sequence, i.e., $j = 1$. Therefore, $b = w_{1}^{i + 1}$, a contradiction with the assumptions of the claim. Thus, $W$ is an atomic saturable sequence.

Now we will show that $W$ is a design. Consider any MFE (saturated) structure $S$ for $W$. Since $W$ is atomic saturable, $a$ is paired with $b$ in $S$. By Claim~\ref{l:atomic1}, $w^{1}\cdots w^{t}$ is a design. It follows that $W$ is a design as well.
\end{dproof}

To prove the sufficiency of the degree condition, consider the following algorithm, which takes as input a saturated structure $S$ with $\MaxDeg{S}\le 2c$, and returns a design $w$ for $S$:
\begin{itemize}
\item 
  Let $\{[l_{i} ,r_{i}]\}_{i = 1 }^{d}$ be the children of the root. Assign to each $w_{l_{i}},w_{r_{i}}$ complementary bases such that $\forall 1\le i < j\le d:\ w_{l_{i}}\neq w_{l_{j}}$.
\item 
  While there exists an unprocessed internal node $[a,b]$ whose parent has been processed  (if there is no such node, stop and return $w$). Let $\{[l_{i} ,r_{i}]\}_{i = 1 }^{d}$ be the children of $[a,b]$. Assign to each $w_{l_{i}},w_{r_{i}}$ complementary bases such that $\forall 1\le i \le d:\ w_{l_{i}}\neq w_{b}$ and $\forall 1\le i < j\le d:\ w_{l_{i}}\neq w_{l_{j}}$.
\end{itemize}

Note that since the alphabet contains $c$ pairs of complementary bases, the assignment at each step of the algorithm is possible. We will show that the returned sequence $w$ is a design for $S$. We will show by tree induction on the size subtrees that $w_{i}\cdots w_{j}$ is an atomic saturable design for every internal node $[i,j]$. It is easy to check that this is satisfied at the leaves. Consider an internal node $u$. By the induction hypothesis, sequences for each child subtree of $u$ are atomic saturable designs. Furthermore, by the choice of bases at children nodes of $u$, all assumptions of Claim~\ref{l:atomic2} are satisfied, hence, the sequence for node $u$ is also an atomic saturable design. The claim holds. Finally, we can apply Claim~\ref{l:atomic1} at the root, which yields that $w$ is a design.
\end{dproof}

\begin{cor}[Result \Result{2}]\label{t:two-letter}
  A structure $S$ is $\Sigma_{1,0}$-designable if and only if it does not contain any base pairs, or it is saturated and $\MaxDeg{S}\le 2$.
\end{cor}

\begin{dproof}
  If $S$ contains a base pair and an unpaired position, then it can be easily checked that $S$ is not $\Sigma_{1,0}$-designable.   
Hence, any $\Sigma_{1,0}$-designable structure is either empty, and trivially designable using a single letter, or saturated. In the latter case, by Theorem~\ref{t:result4}, we know that designable structures are exactly those that are saturated, and such that $\MaxDeg{S}\le 2$. The claim follows.  
\end{dproof}

\begin{cor}[Result \Result{3}]\label{t:three-letter}
  A structure $S$ is $\Sigma_{1,1}$-designable if and only if $$\MaxDeg{S}\le 2.$$
\end{cor}

\begin{dproof}
  First, suppose $S$ is $\Sigma_{1,1}$-designable and let $w$ be a design for $S$. 
Then $\Paired{w,S}$ is a design for the paired restriction $\Paired{S}$ of $S$. Since $\Paired{S}$ is saturated, $\Paired{w,S}$ is over alphabet $\Sigma_{1,0} \subset \Sigma_{1,1}$, and by Theorem~\ref{t:result4}, $D(\Paired{S})\le 2$. Hence, $\MaxDeg{S} = \MaxDeg{\Paired{S}}\le 2$.

  Conversely, suppose that $\MaxDeg{S}\le 2$. Construct a design for $S$ as follows. Since $\Paired{S}$ is saturated, by Theorem~\ref{t:result4}, there is a design $\bar w$ for $\Paired{S}$ over $\Sigma_{1,0} \subset \Sigma_{1,1}$. Construct $w$ from $\bar w$ by inserting the base without a complementary base at every unpaired position of $S$. Let $S'$ be an MFE structure for $w$. Obviously, all unpaired positions in $S$ are also unpaired in $S'$. We must have $\Paired{S'} = \Paired{S}$, otherwise we have an alternative structure for $\bar w$, a contradiction. Hence, $S' = S$, i.e., $w$ is a design for $S$.
\end{dproof}

Result \Result{4} follows immediately from Theorem~\ref{t:result4} by taking $c = 2$.

\begin{lem}[Result \Result{5}]\label{th:forbidden}
  Any structure that contains $\MotifA$ or $\MotifB$ is not $\Sigma_{2,0}$-designable.
\end{lem}

\begin{dproof}
  Assume that $S$ is $\Sigma_{2,0}$-designable and let $w$ be a design for $S$. Then $\Paired{w,S}$ is a design for $\Paired{S}$. Since $\Paired{S}$ is saturated, by Theorem~\ref{t:result4}, $\MaxDeg{S} = \MaxDeg{\Paired{S}}\le 4$, hence, $S$ cannot contain motif $\MotifA $.

Now, assume that $S$ contain motif $\MotifB $ appearing at node $[a,b]$ of $T_{S}$. Let $\{[l_{i},r_{i}]\}_{i = 1 }^{3}$ be some paired children of $[a,b]$ and the node $[a,b]$ if $[a,b]$ is not the root, and $[u,u]$ an unpaired child of $[a,b]$. Let $L_{i} = l_{i}$ and $R_{i} = r_{i}$ if $[l_{i},r_{i}]$ is a child of $[a,b]$, and $L_{i} = r_{i}$ and $R_{i} = l_{i}$ if it is $[a,b]$. If among bases $w_{L_{1}},\dots,w_{L_{3}}$ there is a pair of repeated letters, then we can construct an alternative MFE structure for $w$ (see the first paragraph in the proof of Theorem~\ref{t:result4}). Assume that these three bases are different. Then for some $i = 1,2,3$, $w_{u}$ equals either $w_{l_{i}}$ or $w_{r_{i}}$, say it equals $w_{l_{i}}$. Then $S\setminus \{(l_{i},r_{i})\} \cup \{(u,r_{i})\} $ is an MFE structure for $S$, a contradiction with the assumption that $w$ is a design for $S$.
%
\end{dproof}

\begin{figure}
{\centering
\includegraphics[width=\textwidth]{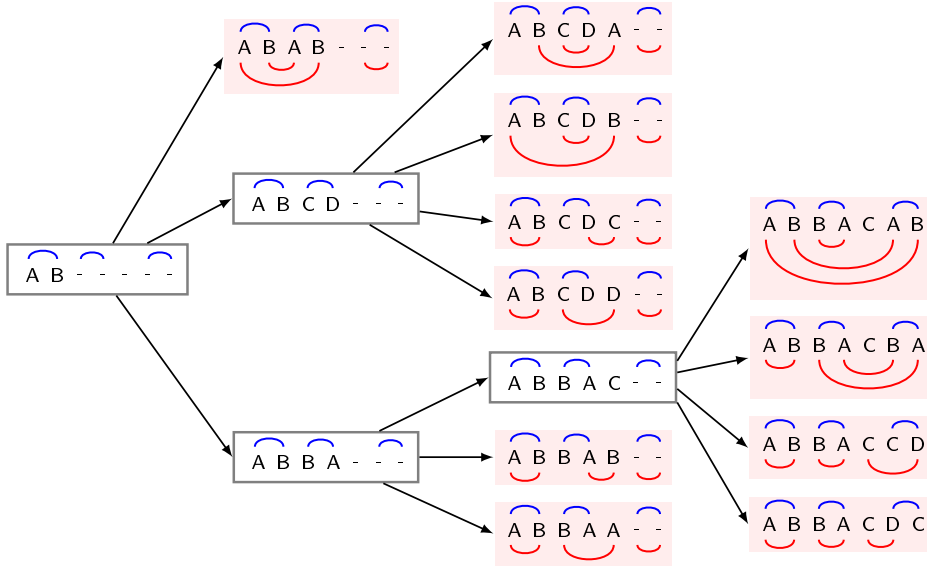}
\\}

\caption{Exhaustive search and systematic counter-examples for the design of $\MotifB$. 
${\sf A}-{\sf B}$ and ${\sf C}-{\sf D}$ respectively represent the first and second pairs of letters found in the design in prefix order, allowing the factorization of trivial symmetries.}
\end{figure}

\TODOJan{Add argument for generalizing to energy models (even with $\Gb\Ub$ base-pairs, as the base-pairing induce a bipartite graph, therefore ...)}

\TODOJan{Fix redundant(?) statement}
\begin{thm}[Result \Result{6}]
  \label{t:result6}
  If the tree representation of a structure $S$ admits a separated coloring then $S$ is $\Sigma_{2,0}$-designable.
\end{thm}
\begin{dproof}
Given a sequence $w$, we define the \Def{level} $\Level{i}$ of position $i$ as $\Level{i} = |\Prefix{i}{w}|_{\Gb}-|\Prefix{i}{w}|_{\Cb}$. 

\begin{clm}
  \label{lem:equalheight}
  Consider any structure $S$ compatible with sequence $w$ that contains some $\Ab-\Ub$ base pair between positions at different levels, then there exists a position $\Gb$ or $\Cb$ that is left unpaired in $S$.
\end{clm}

\begin{dproof}
  Assume that the $\Ab-\Ub$ base pair occurs at position $(a,b)$, and note that the bases of the substring $\Substring{a+1}{b-1}{w}$ can only base pair among themselves without introducing crossings. We will show that $\Gb$'s and $\Cb$'s are not balanced in this substring. Since $w_{b}\in\{\Ab,\Ub\}$, $\Level{b}=\Level{b-1}$. Hence, by the definition of $\LevelFun$, we have that 
  $$|\Substring{a+1}{b-1}{w}|_{\Gb}-|\Substring{a+1}{b-1}{w}|_{\Cb}= \Level{b-1}-\Level{a} = \Level{b}-\Level{a}\neq 0\,.$$
  Therefore, at least one $\Gb $ or $\Cb $ in the substring remains unpaired in this structure.
\end{dproof}

  Consider a separated coloring of the tree representation of $S$. We will use this coloring to construct a design $w$ for $S$, by specifying a nucleotide at each position of $w$. First, for each unpaired position $i$, set $w_{i} = \Ab $. 
Second, apply a modified version of the algorithm described in the proof of the Theorem
~\ref{t:result4} to set the bases of paired positions in which black nodes are assigned to base pair $\Gb - \Cb $, white nodes to $\Cb - \Gb $ and grey nodes to $\Ab - \Ub $ or $\Ub - \Ab $. The algorithm ignores unpaired nodes in the tree representation of $S$. Since the coloring is proper such assignment is always possible at every step of the algorithm. We claim that for any node $[i,j]$ (paired or unpaired), the level of position $i$ is the same as the level of the node $[i,j]$. To verify this, observe that  the substring of $w$ corresponding to any subtree has the same number of $\Gb $'s and $\Cb $'s. Hence, for any node $[i,j]$, the level of position $i$ depends only on nodes on the path from this node to the root. It is easy to check that the level of $i$ is equal to the level of the node. Note that if $[i,j]$ is a grey node then the level of position $j$ is the same as the level of $i$, i.e., the same as the level of $[i,j]$.

We will show that the constructed $w$ is a design for $S$. Since all $\Cb $'s and $\Ub $'s of $w$ are paired in $S$, $S$ is an MFE structure for $w$. We need to show that it is the only MFE structure for $w$. Consider an MFE structure $S'$ for $w$ different from $S$. Since $w$ has the same number of $\Gb $'s and $\Cb $'s, $S'$ must pair all $\Gb $'s, $\Cb $'s and $\Ub $'s of $w$. We will show that all unpaired positions in $S$ are also unpaired in $S'$. Assume to the contrary that position $i$ is unpaired in $S$, but it is paired to $j$ in $S'$. We must have $w_{i} = \Ab $ and $w_{j} = \Ub $. Since the coloring is separated, the unpaired node $[i,i]$ has a different level than the grey node containing $j$, and hence, the level of $i$ is different from the level of $j$. It follows by Claim~\ref{lem:equalheight} that some $\Gb $ or $\Cb $ is unpaired in $S'$, a contradiction.
Consider the paired restrictions of $S$, $S'$ and $w$.
Both $\Paired{S}$ and $\Paired{S'}$ are saturated and compatible with $\Paired{w,S}$ and they are different since $S$ and $S'$ are different and agree on the unpaired positions. Furthermore, $\Paired{w,S}$ can be produced by the algorithm described in the proof of Theorem
~\ref{t:result4} for the input structure $\Paired{S}$, and hence, by Theorem~\ref{t:result4}, $\Paired{w,S}$ is a design for $\Paired{S}$, which contradicts the existence of $\Paired{S'}$. Hence, $w$ is a design for $S$.
\end{dproof}


\begin{figure}
{\centering \includegraphics{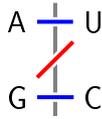}\\}

  \caption{The compatibility graphs of the Watson-Crick (blue edges) and the Nussinov-Jacobson (blue and red edges) energy models are bipartite. }
  \label{fig:bipartiteEM}
\end{figure}

Next, we show the closure of the set of designable structures under the $k$-stutter operation. To that purpose, we introduce the \Def{compatibility graph} of an energy model $\Model$, whose vertices are the four nucleotides $\{\Ab,\Cb,\Gb,\Ub\}$, and whose edges correspond to valid base-pairs in $\Model$, i.e. having finitely-valued contributions.

\begin{defn}[Bipartite energy model]
An energy model $\Model$ is \Def{bipartite}
if and only if its compatibility graph is bipartite.
\end{defn}

The Watson-Crick energy model $\WM$ and the Nussinov-Jacobson energy model
$\WMe$ are bipartite, as can be seen in Figure~\ref{fig:bipartiteEM}.

\begin{thm}[Result  \Result{7}]\label{t:t7}
For any bipartite energy model $\Model$ and any energy
distance $\Delta$, if $w$ is a $\Delta$-design for a structure $S$, then for any integer $k\ge 1$, $w^{[k]}$ is also a
$\Delta$-design for $S^{[k]}$.
\end{thm}

\begin{dproof}
%

  Consider a designable structure $S$ and let $w = w_{1}\cdots w_{n}$
  be a design for $S$. We will show that
  $w^{[k]}$ is a design for $S^{[k]}$. Let us use the $i^{\text{\,th}}$ \Def{block} in $S^{[k]}$ (resp. $w^{[k]}$) as a shorthand for the subset $[1+i\cdot k, 1+(i+1) \cdot k )$ of its positions. Note that the positions involved in the $i^{\text{\,th}}$ block of $S^{[k]}$ correspond to the $i^{\text{\,th}}$ position in $S$ (resp. $w$).

  Consider a valid structure $S'\neq S^{[k]}$ for $w^{[k]}$. Define an \Def{interaction
    multigraph} $I(S') = (V_{I(S')},E_{I(S')})$ of $S'$ as
  follows: the vertex set $V_{I(S')}$ is the set of positions $\{1,\dots,n\}$ in $w$, 
  and there are as many edges between $i$ and $j$ in
  $I(S')$ as there are base-pairs in $S'$ between the $i^{\text{\,th}}$ and $j^{\text{\,th}}$ blocks. Clearly, $I(S')$ is a multigraph of maximal degree $k$. Moreover any edge between the $i^{\text{\,th}}$ and $j^{\text{\,th}}$ block in $I(S')$ corresponds to a valid base-pair $(i,j)$ for $w$. Therefore the sequence of nucleotides read along any path in $I(S')$ must constitute a valid path in the compatibility graph. Since the energy model is bipartite, then any cyclic path cannot have odd length, and so $I(S')$ is also bipartite. 
  
  Since $I(S')$ is a bipartite multigraph of maximal degree $k$, then by K\"onig's theorem \cite{Konig1916} it is $k$ edge-colorable (see \protect{\cite[page 52]{Wilson2002}} for an English version of the proof). In other words, we can color the base-pairs of $S'$, using less than $k$ colors, such
  that each block in $S'$ is involved in at most one base-pair of each color. 
  Therefore we can partition the base-pairs of $S'$ into $k$ structures $S'_1, S'_2, \cdots , S'_{k}$ that are compatible with $w$. Note that the base-pairs of $S'_i$ are pairwise non-crossing since $S'$ itself is non-crossing.
  
  The sequence $w$ is a design for $S$, thus one has
  $$\E{\mathcal{W}}{w}{S'_i} \geq \E{\mathcal{W}}{w}{S},   \text{ for every } 1 \leq i <k.$$
   Moreover, one has $S'\neq S^{[k]}$ so there exists a structure $S'_j$ such that $S'_j \neq S$, and therefore $\E{\mathcal{W}}{w}{S'_j} \geq \E{\mathcal{W}}{w}{S} + \Delta$. It follows that
$$\E{\mathcal{W}}{w^{[k]}}{S'} =\sum_{i=0}^{k-1} \E{\mathcal{W}}{w}{S'_i} \geq k \cdot \E{\mathcal{W}}{w}{S} +\Delta = \E{\mathcal{W}}{w^{[k]}}{S^{[k]}}+ \Delta.$$
 
 We conclude that $S^{[k]}$ is the sole MFE structure for $w^{[k]}$, and has energy at least $\Delta$ less than its foremost competitor, so $w^{[k]}$ is a $\Delta$-design for $S^{[k]}$.
  
  \end{dproof}
  
  Result \Result{7} immediately follows from Theorem~\ref{t:t7}, by reminding the bipartite nature of the Watson-Crick energy model.

\begin{thm}[Result \Result{8}]
Each structure $S$ without $\MotifA$ and $\MotifB$ can be transformed into a $\Sigma_{2,0}$-designable structure $S'$  by inflating a subset of its base pairs (at most one per band). Furthermore, this transformation can be done in $\Theta(n)$ time.
\end{thm}

\begin{dproof}
We start with the greedy coloring of $T_{S}$. Since $S$ does not contain $\MotifA$ and $\MotifB$, it is a proper coloring and there is no node having both a grey child and an unpaired child. We will insert base pairs within $S$ so that the grey nodes and any unpaired node end up at levels of different parities. If the root has a grey child, assign even parity to the grey nodes, otherwise (if the root has an unpaired child, or no grey and no unpaired children), assign even parity to the unpaired nodes.

Now we proceed from the children of the root towards leaves adjusting parity level for grey and unpaired nodes to keep one type even and the other one odd. We repeatedly apply the following simple operation on  $T_{S}$:
If the node $N$ does not match its intended parity level. Denote $N_P$ the parent of $N$ ($N_P$ is not the root as all children of the root already have the correct parity level) and $N_{PP}$ the parent of $N_P$. Insert a new paired node $N_N$ between $N_{PP}$ and $N_P$, assign it with the color of $N_P$, and apply the greedy algorithm on $N_N$. Observe that $N_P$ always takes either black or white color changing the parity level of all its descendants (including $N$). Note that the children of $N_P$ may get recolored, we can even get one more grey child but after this operation the parity levels of all children of $N$ are correct and we do not change parity levels outside the subtree rooted at $N$. After fixing all nodes, we get a separated proper coloring (which is actually the greedy coloring) of $T_{S'}$. Hence, by Theorem~\ref{t:result6}, $S'$ is designable. Figure~\ref{fig:inflation} illustrates this process.
\end{dproof}

\subsection{Nussinov-Jacobson energy model $\cal N$}
\label{sec:weight-wats-crick}

We will show that, for a value of $\Delta\in (0,\min (|\alpha |,|\beta |)]$ our results for the Watson-Crick model transpose to the more general Nussinov-Jacobson  model.

First, we will establish that \Result{1}, \Result{2}, \Result{3} and \Result{7} hold in the $\cal N$ model. 

\Result{1} concerns an alphabet that does not allow base pairs to occur, so their weighting is unconsequential. 
\Result{2} is equally trivial since the uniform weighting of every occurrence of a base pair type does not affect the relative order of structures. 

\Result{3} requires a clarification, as the introduction of two partners $\Ab$ and $\Gb$ for $\Ub$ somehow assign an unambiguous semantics to $\Sigma_{1,1}$ notation, leading to a disjunctive discussion. Let us assume that $\Sigma_{1,1}$ represents $\{\Ab,\Cb,\Gb\}$ (resp. $\{\Ab,\Cb,\Ub\}$), then the argument used for \Result{2} holds, since $\Ab$ (resp. $\Cb$) cannot form base pairs. 
\Result{7} is a direct corollary of Theorem~\ref{t:t7} which is applicable for any bipartite energy model. Since the Nussinov-Jacobson model is bipartite, as shown in Figure~\ref{fig:bipartiteEM}, then \Result{7} also holds in the $\cal N$ model.

\begin{defn}
  Let $X\subseteq \{\Cb ,\Gb, \Ab ,\Ub \} $.
  A design $w$ for a structure $S$ is \Def{$X$-unpaired} if and only if the bases of $w$ found at unpaired positions in $S$ belong to $X$.
  If $X = \{b\} $ is a singleton, the notation is shortened to \Def{$b$-unpaired}. 
\end{defn}

Let $n_{\Gb\Cb }(w,S)$ (resp. $n_{\Ab\Ub }(w,S)$, $n_{\Gb\Ub }(w,S)$) be the number of $\Gb -\Cb $ (resp. $\Ab -\Ub $, $\Gb -\Ub $) base pairs of $S$ on $w$. Note that $$\E{\WM }{w}{S} = -\nGC{w}{S} - \nAU{w}{S}$$ and 
$$\E{\WMe }{w}{S} = \alpha. \nGC{w}{S} + \beta.\nAU{w}{S} + \gamma.\nGU{w}{S}.$$ 





\begin{prop}\label{prop:extended-implication}
  Let $\Delta_{\WM} = 1 $ and $0 < \Delta_\WMe \le \min (|\alpha |,|\beta |)$, if a structure is $\Ab $-unpaired and $(\WM ,\Sigma_{2,0} ,\Delta_\WM )$-designable then it is also $(\WMe ,\Sigma_{2,0} ,\Delta_\WMe )$-designable.
\end{prop}

\begin{dproof}
  Let $w$ be an $\Ab $-unpaired $(\WM ,\Sigma_{2,0} ,\Delta_{\WM} )$-design for $S$. Since in $\WM $ there are no $\Gb - \Ub $ base pairs, we have
$$
\nGC{w}{S} = |w|_\Gb = |w|_\Cb\,,
\qquad 
\nAU{w}{S} = |w|_\Ub\,,
$$
and for any other structure $S'\in \CompSec{|w|}$, 
$$\nGC{w}{S'} \le \nGC{w}{S} \qquad \nAU{w}{S'} \le \nAU{w}{S}\,,$$
with at least one of the inequalities being strict. 

We will show that $w$ is also a $(\WMe ,\Sigma_{2,0} ,\Deltae )$-design for $S$. Consider any alternative structure $S'\in \CompSec{|w|}$. If $\nGU{w}{S'} = 0$, then
\begin{align*}
  \E{\WMe }{w}{S'} &= \alpha\,\nGC{w}{S'} + \beta\,\nAU{w}{S'}\\
  &\ge\,\alpha\nGC{w}{S} + \beta\,\nAU{w}{S} + \min \{|\alpha |,|\beta |\} 
 \ge \E{\WMe}{w}{S} + \Deltae\,,
\end{align*}
as required. Otherwise ($\nGU{w}{S'} > 0$), first observe that
\begin{equation}\label{eq:counts}
  \nGC{w}{S'} + \nGU{w}{S'}\le |w|_{\Gb}
  \quad \text{and}\quad
  \nAU{w}{S'} + \nGU{w}{S'}\le |w|_{\Ub} \,.
\end{equation}
Now, we have
\begin{align*}
  \E{\WMe }{w}{S'} &= \alpha\,\nGC{w}{S'} + \beta\,\nAU{w}{S'} + \gamma\,\nGU{w}{S'}\\
  &\underset{\scriptsize (\ref{eq:counts})}{\ge} 
  \alpha\,|w|_{\Gb } - \alpha\,\nGU{w}{S'} + \beta\,|w|_{\Ub } - \beta\,\nGU{w}{S'} + \gamma\,\nGU{w}{S'}\\
  & = 
  \alpha\,|w|_{\Gb } + \beta\,|w|_{\Ub } + (\gamma - \alpha - \beta )\nGU{w}{S'}\\
 &\ge \E{\WMe}{w}{S} + \gamma - \alpha - \beta. 
 \end{align*}
Moreover, since $\alpha,\beta,\gamma<0$ and $\alpha,\beta<\gamma$, then $$\gamma - \alpha - \beta\ge\max(|\alpha|,|\beta|) \ge \min(|\alpha|,|\beta|) \ge \Deltae.$$
We conclude that 
\begin{align*}
 \E{\WMe }{w}{S'} &\ge \E{\WMe}{w}{S} + \Deltae\,
\end{align*}
as required.
\end{dproof}

\begin{cor} Results 
  \Result{4}, \Result{5}, \Result{6} and \Result{8} hold in any Nussinov-Jacobson energy model $\cal N$ with $0 < \Deltae \le \min (|\alpha |,|\beta |)$.
\end{cor}

\begin{dproof}
The validity of Results \Result{6} and \Result{8} in $\cal N$ follows directly from a close inspection of the constructive proofs in the Watson-Crick energy model, both establishing the existence of $\Ab $-unpaired designs. Proposition~\ref{prop:extended-implication} therefore applies, and extends the validity of those designs to 
 any suitable $\Deltae$.

\Result{5} follows from the fact that, in the proof of Theorem~\ref{th:forbidden}, our counterexamples 'locally' trade one base pair in $S$ for another in the alternative structure, and that the two base pairs are of the same type. Therefore, both structures have the same energy in the Nussinov-Jacobson energy model.

From this, we conclude on the validity of \Result{4}. Indeed, any failure to the degree condition $D(S)\le 4$ implies the existence of $\MotifA$ in $S$, {\it i.e.} such structures are undesignable and the degree condition is therefore necessary.

\end{dproof}






\section{Conclusion, discussion and perspectives}
\label{sec:conclusion}

In this work, we introduced the {\em Combinatorial RNA Design problem}, a {\em minimal} instance of the RNA design problem which aims at finding a sequence that admits the target structure as its unique base pair maximizing structure using $\Ab-\Ub$ and $\Gb-\Cb$ base pairs. First, we provided complete characterizations for the structures that can be designed using restricted alphabets. Then we considered the RNA design under a four-letters alphabet, and provide a complete characterization of designable saturated structures, i.e., free of unpaired positions. Turning to those target structures that contain unpaired positions, we provided partial characterizations for classes of designable/undesignable structures, and showed that the set of designable structures is closed under the stutter operation. Finally, we introduced structure-approximating version of the problem and, assuming that the input structure avoids two motifs, provided a structure approximating algorithm of ratio $2$ for general structures. 
We showed that our results also hold in the more realistic Nussinov-Jacobson energy model, which allows $\Gb-\Ub$ base pairs to occur, and associates arbitrary negative free-energy to each base pair type ($\Gb-\Ub$ being the weakest).

An important question that is left open by this work is the computational complexity of the RNA design problem. Schnall-Levin~\emph{et al.}~\cite{Schnall-Levin2008} established the {\sf NP}-hardness of a more general problem, called the inverse Viterbi algorithm, which takes as input a stochastic grammar (representing the energy model) and a targeted parse tree (representing the structure), and outputs a sequence (design) whose most probable parsing should match the target. However this result does not settle the complexity of the RNA design, essentially because the proposed reduction relies critically on an encoding of 3-SAT instances within the input grammar. While the hypothetical {\em perfect} grammar/energy model for RNA folding probably differs from the currently accepted Turner model, it should ultimately reflect the laws of physics and should certainly not depend on the instance. As the reduction~\cite{Schnall-Levin2008} requires a different grammar (i.e., energy model) for each instance, it does not seem easily adaptable into a proof that holds for a fixed energy model. Consequently, despite two decades of work on the subject, the computational tractability of RNA design is still open, either in its general instance and in our purely combinatorial version.

In our opinion, this exceptional resistance of the RNA design problem to any attempt so far at characterizing its computational complexity can be attributed to two main reasons:
\begin{itemize}
\item The inverse nature of the problem: While polynomial, the direct computation MFE folding for an RNA requires dynamic programming, and runs in $\Theta(n^3)$ time (up to polylogarithmic factors, see ~\cite{Zakov2011} for a complete state-of-the-art). Unfortunately, the {\em optimal-substructure} property of the direct problem does not transpose to the inverse problem. 

Therefore, solving the RNA design problem somehow requires inverting a non-trivial -- yet polynomially computable -- function. It is tempting here to draw a parallel with some areas of cryptography, where multiple protocols are based on a -- sometimes difficult to establish -- disymmetry between the complexities of the direct and inverse computation.
\item The intricacies of the objective function: Current state-of-the-art implementations of MFE folding prediction algorithms rely on a sophisticated energy model, the Turner model~\cite{Turner2010}. This model associates energy contributions to as much as 24~000 different types of structure/sequence motifs, and vastly increases the complexity the characterization of the space, and energies, of competing structures.

  On the other hand, oversimplified statements for the problem, as would result from a relaxation of the uniqueness condition, can be trivially solved in linear time. Such problems are not only largely unrealistic from a biological perspective, but they also probably do not retain the potential difficulty of the general problem.
\end{itemize} 

Besides complexity issues, natural extensions of this work may include the consideration of more sophisticated energy  models such as those based on stacking pairs or, ultimately, to the full Turner energy model~\cite{Turner2010}).
One could also consider incorporating additional constraints, expressed as the presence/avoidance of motifs~\cite{Zhou2013}, $\Gb{}\Cb$-content~\cite{Reinharz2013}\ldots{} or the design under other objectives, such as the Boltzmann probability~\cite{Zadeh2011}.
In the Nussinov-Jacobson model, our result could be completed by the consideration of more liberal values for $\Delta$, although it should be noted that considering larger such values would only gradually deplete the sets of designable structures until it becomes empty when the conditions of Theorem~\ref{eq:condDelta} are met.
More precise bounds for the ratio of the structure-approximating could also be established. Finally, the structure-approximation problem could be revisited in an optimization setting, in which one would attempt to minimize the number of modifications made to the structure, so that a given structure becomes designable (or, more modestly, belongs to an identified class of designable structures). We plan to address some of these questions in future works.


\bibliographystyle{spmpsci}      
\bibliography{biblio}   

\end{document}